\documentstyle [12pt] {article}
\title  {$C_{E}PT$ SYMMETRY OF THE SIMPLE ECOLOGICAL  DYNAMICAL EQUATIONS}

\author{Vladan Pankovi\'c$^\ast$ , Rade Glavatovi\'c$^\sharp$ ,Milan Predojevi\'c $^\ast$ \\
$^\ast$Department of Physics , Faculty of Science, \\ 21000 Novi
Sad , Trg Dositeja Obradovi\'ca 4. ,
Serbia and Montenegro , \\
$^\sharp$Military Medical Academy \\11000 Beograd , Crnotravska 17
, Serbia and Montenegro \\
vladanp@gimnazija-indjija.edu.yu}
\date {}
\begin {document}
\maketitle

\vspace {1.5cm}

\begin {abstract}

It is shown that all simple ecological, i.e. population dynamical
equations (unlimited exponential population growth (or decrease)
dynamics, logistic or  Verhulst equation, usual and generalized
Lotka-Volterra equations)  hold a symmetry, called $C_{E}PT$
symmetry. Namely, all simple ecological dynamical equations are
invariant (symmetric) in respect to successive application of the
time reversal transformation - $T$, space coordinates reversal or
parity transformation - $P$, and predator-prey reversal
transformation - $C_{E}$ that changes preys in the predators or
pure (healthy) in the impure (fatal) environment, and vice versa.
It is deeply conceptually analogous to remarkable $CPT$ symmetry
of the fundamental physical dynamical equations. Further, it is
shown that by more accurate, "microscopic" analysis,  given
$C_{E}PT$ symmetry becomes explicitly broken.

\end{abstract}

\section {Introduction. $CPT$ theorem in the fundamental physical dynamical equations}

As it is well-known [1], [2] that $CPT$ symmetry theorem
represents one of the most important theorem in the fundamental
physics, more precisely relativistic quantum mechanic or local
quantum field theory. Within less fundamental, classical physics
or nonrelativistic quantum mechanics, $CPT$ theorem cannot be
proved but some its consequences exist even in these less
fundamental physical theories.

$CPT$ theorem is  based and can be proved by some of the most
fundamental concepts of the local quantum field theory. First one
represents the dynamical, continuous Lorentz invariance
(symmetry). It, simply speaking, admits only local (realizable
luminal, i.e. by speed of light) and forbids any nonlocal
(superluminal, i.e. faster than light) dynamical interaction. Or
the final  (in the future) physical dynamical state represents the
result of the non-instantaneous dynamical evolution of the initial
(in the past) physical dynamical state. Second one represents the
quantum field theory postulate that an elementary particle can
have either integer spin (spin  that equals even number of $\frac
{\hbar}{2}$ where $\hbar$ represents reduced Planck's constant) or
half-integer spin (spin  that equals odd number of $\frac
{\hbar}{2}$).

$CPT$ theorem refers on the dynamical discrete symmetries and  it
states the following. All fundamental physical dynamical equations
and theirs solutions are invariant (symmetric) in respect to
successive  application of the following three discrete
transformations: time reversal transformation - $T$, space
coordinate reversal or parity transformation - $P$, and charge
conjugation transformation - $C$. Simply speaking, $T$ changes
time moment $t$ in $-t$ and initial in the final conditions, $P$
changes space coordinates $x$, $y$, $z$  in $-x$, $-y$, $-z$,
while $C$ changes any charge parameter $c$ in $-c$,or more
generally, all particles in antiparticles, and vice versa.

On the basis of $CPT$ theorem significant spin-statistic theorem
can be proved. Also, on the basis of $CPT$ theorem it follows that
particle and antiparticle masses and  lifetimes must be
equivalent, while theirs electrical charges and magnetic moments
must be opposite, i.e. must have opposite signature.

Neither $T$ nor $P$ nor $C$ represents any real dynamical motion.
In this sense  all three given transformations are abstract.
However, all of them, or at least $P$ and $T$, can be relatively
simply formally demonstrated. $P$ can be simply particularly
demonstrated by a mirror, i.e. mirror transformation of the space.
Also, $T$ can be simply demonstrated by a video projector when it
does a play back backward (corresponding to negative, i.e.
reversal time direction). (Of course, when a video projector does
play back forward it corresponds to usual, i.e. positive time
direction. So if  a physical process recorded by video camera and
presented by video projector is time reversible or T-symmetric (as
it is motion of a linear harmonic oscillator, eg. undamped
pendulum), then both, forward and backward, play backs are
empirically "natural". But if a physical process recorded by video
camera and presented by video projector is time irreversible or
$T$-asymmetric (as it is diffusion of a drop of the ink in a glass
of the water) then only one, forward play back is empirically
"natural", while other, backward play back projection, is
empirically "strange" or "impossible".

In this work it will be shown that all simple ecological, i.e.
population dynamical equations (unlimited exponential population
growth (or decrease) dynamics, logistic or  Verhulst equation,
usual and generalized Lotka-Volterra equations) [3]-[6] hold a
symmetry, called $C_{E}PT$. Namely  all simple ecological
dynamical equations are invariant (symmetric) in respect to
simultaneous application of $T$, $P$, and predator-prey reversal
transformation - $C_{E}$, that changes preys in the predators or
pure (healthy) in the impure (fatal) environment, and vice versa.
It is deeply conceptually analogous to $CPT$ symmetry of the
fundamental physical dynamical equations. Further, it will be
shown that by more accurate, "microscopic" analysis, $C_{E}PT$
symmetry  become explicitly broken.

\section {C$_{E}$PT theorem in the simple ecological dynamical equations}

Consider usual Lotka-Volterra, i.e. predator- prey differential
equations system with significant applications in the ecology,
biology, medicine (epidemiology, neurology) etc. [3]-[6]. Namely,
as it is well-known, given system (including its different
generalizations) represents one of the most simplest and most
typical models of the ecological, i.e. population dynamics. This
system is given by the following  two nonlinear differential
equations of the first order
\begin {equation}
\frac {dp}{dt} = {\it a}p - {\it b}pq
\end {equation}
\begin {equation}
\frac {dq}{dt} = -{\it c}q + {\it d}pq
\end {equation}

Here, mathematically $p$, $q$, represent the real, positive
variables  that depends of the time, $t$, while ${\it a}$, ${\it
b}$, ${\it c}$, ${\it d}$, represent mathematically real, positive
time independent constants, i.e. parameters.

But, ecologically, $p$ represents the  population or number of the
individuals of a biological species of the preys (without physical
dimension). Correspondingly  $q$ represents ecologically the
population or number of the individuals of a biological species of
the predators (without physical dimension). In this way, as it is
well-known, (1), (2) refers on two competitive species living
together. Also, ecologically, ${\it a}$ represents the  birth rate
of the preys, ${\it b}$ - the death rate of the preys, ${\it c}$ -
the death rate of the predators, and ${\it d}$ - the birth rate of
the predators, (all with [s-1] physical dimension). Simply
speaking ${\it a}$, ${\it b}$, ${\it c}$, ${\it d}$ parameters,
more precisely ecological dynamical parameters, express simplified
or phenomenologically the basic dynamical interactions between
predators, preys and environment (food resources, territory,
etc.).

It is necessarily that the following be pointed out. System (1),
(2) describes satisfactorily given population dynamics only then
if $p$ and $q$ are sufficiently large or "macroscopic", i.e. for
\begin {equation}
p \gg 1
\end {equation}
\begin {equation}
q \gg 1
\end {equation}
so that change of $p$ or $q$  for a small (eg. 1, 2  etc.) number
of the corresponding individuals can be considered effectively as
an infinitesimal change. It represents very important
approximation condition. In other words presented simplified form
of the exactly very complex interaction between given two animal
species and environment, corresponding to described population
dynamics (1), (2),  is consistently applicable and can be
correctly applied only then if given "macroscopic" approximation
conditions (4), (5) are  satisfied. Or, according to given
"macroscopic" approximation conditions, more detailed analysis
referring on the small number of the individuals, eg. one
individual, based on (1), (2), is  or can be (mostly) incorrect.
Small values of $p$ and $q$ can be considered consistently only in
the sense of an analytical extension of the solution of (1), (2)
but not in the sense of a more accurate empirical analysis of the
small number of the individuals of the species. For this reason
more detailed (accurate), "microscopic", analysis of the
ecological, i.e. population dynamics of a small number of the
individuals can need significant corrections and changes of (1),
(2).

Thus, ecological, i.e. population dynamics (1), (2), according to
"macroscopic" approximation conditions (3), (4), represents
implicitly a typical statistical  method where populations $p$ and
$q$ are proportional to statistical distributions that evolve
during time. Explicit normalization of $p$ and $q$ to statistical
distributions by some normalization parameters needs, since (1)
and (2) are nonlinear, corresponding normalization of ${\it b}$
and ${\it d}$. In this way in different equivalent forms of the
ecological dynamical system (1), (2) as well as in other
ecological dynamical systems, statistical normalization parameters
without real dynamical sense can appear.

It can be observed that $p$ and $q$ in the usual Lotka-Volterra
equations system (1), (2) as well as  similar simple ecological
dynamical systems (that will be considered later) do not depend of
the space coordinates $x$, $y$, $z$  that characterize predator
and prey species territory (area). (Coordinate dependence of $p$
and $q$ can characterize some more complex ecological dynamical
systems.) For this reason  it can be considered that $p$ and $q$
are effectively space homogeneous and that they are invariant
(symmetric) in respect to parity transformation $P$ (that changes
$x$, $y$, $z$ in $-x$, $-y$, $-z$ respectively). It means that
both equations of the usual Lotka-Volterra equations system (1),
(2) is invariant (symmetric) in respect to $P$ too.

Further, it  can be observed that (1) and (2) have not completely
equivalent form. Namely, the coefficients before $p$ and $pq$ in
(1), ${\it a}$ and ${\it b}$,  are always positive and negative,
while constant before $q$ and $pq$ in (2),  $-{\it c}$ and ${\it
d}$, are oppositely, always negative and positive, respectively.
This in-equivalence or asymmetry causes that $p$ can correspond to
the preys species and that $q$ corresponds to the predators
species exclusively. Or  this non-equivalence forbids that $p$
corresponds to the predators species and that $q$ corresponds to
the preys species.

Define, however, such transformation, called predator-prey
reversal transformation -  $C_{E}$, that changes  ${\it a}$, ${\it
b}$, ${\it c}$, ${\it d}$  in ${\it -a}$, ${\it -b}$, ${\it -c}$,
${\it -d}$ respectively. So ecologically, $C_{E}$ represents a
discrete transformation of the ecological dynamical parameters
only that does not transform pure normalization parameters. Or
generally, $C_{E}$ denotes a change of the direction of the
dynamical interactions between all constituents in the ecological
system.

Obviously $C_{E}$  holds idempotent characteristic $C_{E}^{2} = I
$,where $I$ represents the identity transformation.

Application of $C_{E}$  at (1),( 2) yields
\begin {equation}
\frac {dp}{dt} = -{\it a}p + {\it b}pq
\end {equation}
\begin {equation}
\frac {dq}{dt} =  {\it c}q - {\it d}pq
\end {equation}
or after transposition of the equations,
\begin {equation}
\frac {dq}{dt} =  {\it c}q - {\it d}pq
\end {equation}
\begin {equation}
\frac {dp}{dt} = -{\it a}p + {\it b}pq
\end {equation}

It is not hard to see that (7), (8) represents a typical usual
Lotka-Volterra system but where, formally ecologically, $q$
represents a new preys species population, $p$ - a new predators
species population, ${\it c}$ - the  birth rate of the new preys
species, ${\it d}$ - the death rate of the new preys species,
${\it a}$ - the death rate of the new predators species, and ${\it
b}$ - the birth rate of the new predators species. Simply speaking
$C_{E}$  changes the previous preys species in the new predators
species and the previous predators species in the new preys
species.

In this way $C_{E}$ represents a symmetry transformation of the
set of all usual Lotka-Volterra equations systems (for all values
of the parameters). Or in the domain of the applicability of all
usual Lotka-Volterra equations systems,  it is impossible differ
represents some usual Lotka-Volterra system a real ecological
system or the result of an abstract application of $C_{E}$ at
corresponding real Lotka-Volterra system. But $C_{E}$ does not
represent a symmetry transformation of any concrete usual
Lotka-Volterra equations systems (for concrete values of the
parameters).

Consider now  an especial case
\begin {equation}
{\it b} = {\it d} = 0
\end {equation}
when the usual Lotka-Volterra system (1), (2) turns it the
following system
\begin {equation}
\frac {dp}{dt} = {\it a}p
\end {equation}
\begin {equation}
\frac {dq}{dt} = -{\it c}q
\end {equation}

Obviously, (10), (11) represent a system of two independent
equations (and variables). They  describe unlimited exponential
population growth (increase) and decrease of the population,
respectively, by interaction of the corresponding  species with
its environment. For this reason $p$ species does not represents
preys for $q$ species. But, (10) and (11) have not completely
analogous form since coefficient before $p$ in (10), ${\it a}$, is
positive, while the coefficient before $q$ in (11), ${\it -c}$, is
negative. Ecologically it means that environment positively
stimulates increase of $p$, i.e. first species population. Simply
speaking, environment is pure (healthy) for the first species.
Simultaneously, environment negatively stimulates $q$ i.e. second
species population. Or environment positively stimulates decrease
of $q$. Simply speaking, environment is impure (fatal) for the
second species.

Application of $C_{E}$  at (10),(11) yields
\begin {equation}
\frac {dp}{dt} = -{\it a}p
\end {equation}
\begin {equation}
\frac {dq}{dt} = {\it c}q
\end {equation}
or, after transposition of the equations,
\begin {equation}
\frac {dq}{dt} = {\it c}q
\end {equation}
\begin {equation}
\frac {dp}{dt} = -{\it a}p
\end {equation}

Obviously, (14)(15) represents again a typical system (10)(11) but
now environment is pure (healthy) for $q$ species, while
environment is impure (fatal) for $p$ species. Simply speaking
here  $C_{E}$  changes pure (healthy) in the impure (fatal)
environment and vice versa. It is again in full agreement with
previous general definition that $C_{E}$  denotes a change of the
direction of the dynamical interactions between all constituents
in the ecological system.

Also, it is obvious that $C_{E}$ represents a symmetry
transformation of the set of all (10), (11) equations systems (for
all values of the parameters). Or, in the domain of the
applicability of all (10), (11) equations systems  it is
impossible differ represents some (10), (11) equations system a
real ecological system or the result of an abstract application of
$C_{E}$ CE at corresponding real (10), (11)  equations system .
But , of course , $C_{E}$  does not represent a symmetry
transformation of any concrete (10), (11) equations systems (for
any concrete values of the parameters). It is in full agreement
with previous conclusion on $C_{E}$  as a symmetry transformation
of the set of all  usual Lotka-Volterra systems and an asymmetry
transformation of any concrete usual Lotka-Volterra system.

It is not hard to see (which will not be considered explicitly
here) that same conclusions on the ecological meaning and
characteristics of $C_{E}$  can be done even in case when
unlimited exponential population growth (or decrease) dynamics is
changed by limited exponential population growth (or decrease)
dynamics , i.e. by  logistic or Verhulst population dynamical
equation  [4]-[6]. Namely, in this case, instead of (10), (11)
there are following two equations
\begin {equation}
   \frac {dp}{dt} = {\it a}p \frac {(p_{0} - p)}{p_{0}}
\end {equation}
\begin {equation}
   \frac {dq}{dt} = -{\it c}q \frac {(q_{0} - q)}{q_{0}}
\end {equation}
where $p_{0}$ and $q_{0}$  represent limits of the first and
second species population, while ${\it a}$ and ${it -c}$ represent
corresponding Malthusian parameters. It is very important that the
following be pointed out. In distinction from ${\it a}$ and ${\it
-c}$ that represents real dynamical parameters, $p_{0}$ and
$q_{0}$ do not represent real dynamical parameters but only
normalization parameters. They do not represent any real ecologic
dynamical interaction but express only proportion between
variables and and real numbers of the individuals in corresponding
species. For this reason $C_{E}$  does not any influence at given
normalization parameters.

Now,  apply time reversal T at usual Lotka-Volterra equations
system (1),(2) that yields
\begin {equation}
   \frac {dp}{(-dt)} = {\it a}p - {\it b}pq
\end {equation}
\begin {equation}
   \frac {dq}{(-dt)} = -{\it c}q + {\it d}pq
\end {equation}
which, after simple transformations including transposition of the
equations, becomes equivalent to system (7), (8) representing, as
it has been shown, result of $C_{E}$  application at the same
equations system (1), (2).

Then, on the basis of the previous discussion of the
characteristics of $C_{E}$  transformation of usual Lotka-Volterra
equations system (1), (2) it follows simply that $T$ represents a
symmetry transformation of the set of all usual Lotka-Volterra
equations system  (1), (2) (for all values of the parameters)  and
that, $T$ does not represent a symmetry transformation of any
concrete usual Lotka-Volterra equations system (1), (2) (for any
concrete values of the parameters).

Also, on the same basis, it follows that successive application of
$C_{E}$   and $T$ at any concrete usual Lotka-Volterra equations
system (1), (2) (for any concrete values of the parameters)
represents a symmetry, moreover the identity transformation of
this system. Or, symbolically $C_{E}T = I $.

In this way it is shown that usual Lotka-Volterra system (1), (2)
is invariant (symmetric) in respect to $C_{E}T$, moreover,
$C_{E}PT$  which means that it is invariant (symmetric) in respect
to the successive application of $T$, $P$ and  $C_{E}$.

It is not hard to see  that the equations system (10), (11), i.e.
unlimited exponential population increase and decrease laws, as
well as  system (16), (17), i.e. limited exponential population
increase and decrease laws or logistic, i.e. Verhulst equations,
are, also, $C_{E}T$, moreover, $C_{E}PT$ invariant (symmetric).

Finally, consider  a well-known [4]-[6] generalization of the
Lotka-Volterra equations system
\begin {equation}
\frac {dp_{i}}{dt} = a_{i}p_{i} + \sum_{ij} b_{ij} p_{i}p_{j}
\hspace{1cm} \rm{for} \hspace{1cm} i,j=1,2, …
\end {equation}
where  $p_{i}$ represents corresponding real, time dependent
variable and $a_{i}$ and $b_{ij}$ corresponding real constants for
$i, j= 1, 2, …$. Such system can describe a generalized ecological
dynamics. It is not hard to see that given system is invariant
(symmetric) in respect to $C_{E}T$, moreover, $C_{E}PT$ too.

So, it is shown that all simple ecological dynamical equations,
definable at the "macroscopic" level of the analysis accuracy, are
invariant (symmetric) in respect to $C_{E}PT$, or in respect to
successive application of time reversal, parity and predator-prey
reversal. This statement will be called ecological $C_{E}PT$
theorem.

\section {Explicit breaking of C$_{E}$PT symmetry by "microscopic" analysis}

Now, it will be shown that $C_{E}PT$  symmetry of the  simple
ecological dynamical equations (that satisfy three important
conditions) becomes explicitly broken at the level of a more
detailed (accurate), "microscopic", analysis. For reason of the
simplicity, it will be demonstrated explicitly for unlimited
population growth and decrease dynamics only. But, as it is not
hard to see, the same $C_{E}PT$  symmetry breaking by more
detailed (accurate), "microscopic", analysis, is satisfied for all
other simple ecological dynamical equations.

Consider a simple monosexual biological species without enemies,
i.e. predators that uses unlimitedly environmental resources
(food, territory, etc.). Suppose, firstly, that  population, i.e.
number of the individuals of given species, $p$, is sufficiently
large  so that approximation condition (3) is satisfied.  Then
"macroscopically" definable population dynamics is given by the
expression (10) for pure (healthy) environment or (12) for inpure
(fatal) environment. As it has been discussed, from aspect of
noted "macroscopic" accuracy of the analysis, set of all (10),
(12) equations systems  is $C_{E}$, $P$, $T$ and $C_{E}PT$
symmetric.

Suppose, however, that video camera records population dynamics
with large, "microscopic", resolution (analysis accuracy).
Precisely, suppose that given video camera does not observe
characteristics of the environment but that it  observes some
characteristics of the individuals neglected by previous
"macroscopic" resolution (analysis accuracy). It means that here
approximation condition (3) is unsatisfied which causes that p
cannot represent a continuous time dependent variable so that
dynamical equations (10) and (12) cannot be consistently applied.

For this reason, for "microscopic"description of the population
dynamics, instead of the differential equations (10), (12)
corresponding difference equations
\begin {equation}
\frac {\Delta p_{n+1}}{\Delta t} = {\it a}p_{n} \hspace{1cm} for
\hspace{1cm} n = 1,2,…
\end {equation}
\begin {equation}
   \frac {\Delta  p_{n+1}}{ t} = -{\it a}p_{n}  \hspace{1cm}for \hspace{1cm} n = 1,2,…
\end {equation}
will be suggested which will be called {\it correspondence
condition ( principle)} (in sense that (10), (12) and (21), (22)
have corresponding analogous forms). Here $\Delta t$ represents a
finite time interval of the duration of one dynamical cycle,
$p_{n}$ - natural number of the individuals at the beginning of
$n$-th dynamical cycle, $p_{n+1}$ - natural number of the
individuals at the beginning of $n+1$-th dynamical cycle or at the
end of $n$-th dynamical cycle, while a natural number
\begin {equation}
\Delta p_{n+1} = p_{n+1} - p_{n}
\end {equation}
represents finite difference between population at the end and
beginning of the $n$-th dynamical cycle, for $n = 1,2, …$. Also,
according to the previous discussions, equation (21) refers on the
positive influence of the environment on the species (pure
environment), while equation (22) refers on the negative influence
of the environment on the species (inpure (fatal) environment).

Demand that population value, by "microscopic" analysis, must be a
natural number 1, 2, … etc.  or  eventually  0, represents a quite
natural biological condition, called {\it condition of the
biological discretization (discontinualization) of the
population}.

In equations (21), (22), theirs left hand-sides, representing
change of the population during one dynamical cycle, are
proportional to the (initial) value of the population at the
beginning of this dynamical cycle. In some sense this
"retardation"  represents a "local" characteristic of given
population dynamics or {\it locality condition}. On the contrary,
under a formal supposition that change of the population during
one dynamical cycle is proportional to the (final) value of the
population at the end of this dynamical cycle, this formal
"advance" would represent a "nonlocal" characteristic of such
population dynamics or {\it nonlocality condition}.

It is very important that the following be pointed out. In
distinction from "macroscopic" analysis, here, in "microscopic"
analysis, difference equations (21), (22) do not represent
approximate form of the exact differential equations (10), (12).
Here (21), (22)  represent exact difference equations that
describe ecological dynamics since differential equations (10),
(12) are definitely inapplicable for ecological dynamics
description.

Also, it can be supposed
\begin {equation}
{\it a} = \frac {k}{\Delta t}
\end {equation}
where k represents some number whose meaning will be discussed
later.

According to (24), (21) turns in
\begin {equation}
\frac {\Delta p_{n+1}}{\Delta t} = \frac {k}{ t} p_{n}
\end {equation}
and according to (23) in
\begin {equation}
p_{n+1}-p_{n} = kp_{n}
\end {equation}
for n = 1,2, …  . Finally, it yields
\begin {equation}
p_{n+1} = (k+1) p_{n}    \hspace{1cm} for \hspace{1cm} n = 1,2,
\end {equation}
Since both, $p_{n+1}$ and  $p_{n}$, represent natural numbers for
$n=1,2,…$, then $k+1$ must represent a natural number too. It
means that $k$ can be 0 or any natural number.

For $k = 0$, or according to (24), $ {\it a} = 0$, (27) yields
\begin {equation}
p_{n} = const         \hspace{1cm} for  \hspace{1cm} n = 1,2, …
\end {equation}
which means that here environment does not any influence at any
individual of the species. In this case we can speak on the
neutral environment. But it represents ecologically trivial
situation.

For $k > 0 $, or, according to (24), ${\it a} > 0$, (27) defines
an increasing geometric progression with coefficient $k+1$ greater
than 1. It can be observed that here $k$ can be any natural
number, i.e. that there is no limit for $k$ value.

In this way, for ${\it a} > 0$ and $k > 0$, $\Delta t$ can
consistently represent a finite time interval of the individuals
reproduction, or, time interval of the one reproduction cycle.
Also, here $k$ can consistently represent a natural number that
denotes the additional number of the individuals in respect to one
individual that appear in one reproduction cycle. Or, here, it can
be consistently supposed (within an admitable idealization) that
during one reproduction cycle there is splitting of any individual
(parent) in $k+1$ individuals (descendants) where constant $k$
represents a natural number.

According to (24), (22) turns in
\begin {equation}
\frac {\Delta p_{n+1}}{\Delta t} = - \frac {k}{\Delta t}p_{n}
\end {equation}
and , according to (23), in
\begin {equation}
p_{n+1} - p_{n} = - kp_{n}
\end {equation}
for n = 1,2, …  . Finally, it yields
\begin {equation}
p_{n+1} = (1 - k) p_{n} \hspace{1cm} for \hspace{1cm} n = 1,2, … .
\end {equation}
Since both, $p_{n+1}$ and  $p_{n}$, represent natural numbers for
$n=1,2, …$, then $1 - k$ must represent a natural number too. It
means that $k$ can be 0 or 1. Namely, for $k \neq 0, 1$ population
becomes negative which has none ecological meaning, i.e. which
contradicts to condition of the biological discretization of the
population.

For $k = 0$, or, according to (23), ${\it a} = 0$, it follows
\begin {equation}
p_{n} = const       \hspace{1cm} for \hspace{1cm} n = 1,2, …
\end {equation}
which means that here we have again an ecologically trivial
neutral  environment.

For $k = 1$, or,  according to (23), $| -{\it a} | = \frac
{1}{\Delta t}$ , it follows
\begin {equation}
p_{n+1} = 0  \hspace{1cm} for \hspace{1cm} n = 1,2, … \hspace{1cm}
and \hspace{0.2cm}for \hspace{0.2cm}any \hspace{1cm} p_{1}.
\end {equation}
It means that whole species, for any value of $p_{1}$, die already
during the first dynamical cycle and that $\Delta t$ represents
the life time of any individual. Or, "microscopic" population
dynamics  for negative environmental influence needs that
population (33) represents necessarily  a step function of time
whose initial value at the beginning of the first dynamical cycle
$p_{1}$ can be arbitrary natural number, but whose final value at
the end of the first (and all other) dynamical cycle must be zero
exclusively.  Metaphorically speaking, environmental influence at
given species is here absolutely fatal.

In this way it is shown that, in distinction from set of all
"macroscopic" population dynamics (10), (12), set of all
"microscopic" population dynamics (21)-(24) (based on the
correspondence, biological discretization and locality condition)
is $C_{E}$   asymmetric. Or here change of ${\it a}$ by ${\it -a}$
in (21) does not yield (22), or, change of ${\it a}$ by ${\it -a}$
in (22) does not yield (21). Namely, here positive environmental
influence admits that $k$ can be any natural number, while
negative environmental influence needs that $k$ represents 1
exclusively.

Now, apply $T$ at (21) or (25) for $k > 0$. Practically it means
that here n turns in some m  while  $n+1$ turns in some $m-1$ for
$n = 1,2, … $and $m = 2,3, … $.  It yields
\begin {equation}
-\frac {\Delta p_{m}}{\Delta t} = \frac {k}{\Delta t}p_{m}
\hspace{1cm} for \hspace{1cm} m = 2,
\end {equation}
or
\begin {equation}
\frac {\Delta p_{m}}{\Delta t} = -\frac {k}{\Delta t}p_{m}
 \hspace{1cm} for \hspace{1cm} n = 2, …
\end {equation}
which yields
\begin {equation}
p_{m-1} -  p-{m} = kp_{m}   \hspace{1cm} for \hspace{1cm} n = 1,2,
…
\end {equation}
or, finally,
\begin {equation}
p_{m} = \frac {p_{m-1}}{k+1} \hspace{1cm} for \hspace{1cm} n =1,2,
…
\end {equation}
Obviously, (37) defines a decreasing geometric progression with
coefficient $\frac {1}{k+1}$ smaller than 1.

Obtained geometric progression (37) represents numerically the
inversion of the geometric progression (27)  and vice versa.

Meanwhile, it is obvious that  equation (35) does not satisfy
locality condition but that it satisfies nonlocality condition.
Concretely, change of the population during a cycle depends of the
population at the end of this cycle. For this reason equation (35)
represents neither  the equation   (29) nor equation (25) in which
$k$ is changed by $-k$.

Moreover, a visual demonstration that $T$ applied on (35)  yields
neither (29) nor (25) in which $k$ is changed by $-k$ can be done.
Namely, by time reversal, i.e. by video projector that does a play
back backward, in this case, an observer can see the following.
During  $\Delta t$   it seems that $k+1$ initial individuals
interact mutually which yields only one final individual. In other
words it can seem that during $\Delta t$ one individual eats all
$k$ other individuals, or, that within given biological species
cannibalism appears .

Analogously, apply $T$ at (22) or (29)  for $k = 1$. It yields
\begin {equation}
-\frac {\Delta p_{m}}{\Delta t} = - \frac {k}{\Delta t}p_{m}
\hspace{1cm} for \hspace{1cm} m = 2, …
\end {equation}
or
\begin {equation}
\frac {\Delta p_{m}}{\Delta t} = \frac {k}{\Delta t}p_{m}
\hspace{1cm} for \hspace{1cm} m = 2, …
\end {equation}
which yields
\begin {equation}
p_{m} -  p_{m-1} = kp_{m}     \hspace{1cm} for \hspace{1cm} m = 2,
…
\end {equation}
or, finally,
\begin {equation}
(1 - k) p_{m} = p_{m-1}   \hspace{1cm} for \hspace{1cm} m = 2, …
\end {equation}
i.e. , since k=1,
\begin {equation}
p_{m-1} = 0    \hspace{1cm} for \hspace{1cm} m = 2, … \hspace{1cm}
and \hspace{0.2cm} for \hspace{0.2cm}any \hspace{1cm} p_{M}
\end {equation}
where $M$ represents maximal value of $m$. Obviously, according to
(42), population at the beginning and end of any cycle (except at
the end of the final cycle) must be 0, while population at the end
of the final dynamical cycle can be any natural number.
Mathematically, obtained time dependent population represents a
discrete step function in some degree inverse to step function
(33). But, ecologically, statement that population at the end of
any cycle except the final, must be 0, while population at the end
of the final dynamical cycle can be any natural number is
meaningless. Namely, such statement means in fact that individuals
of given species can be obtained by environment influence only
without theirs parents.

Also, it is obvious that  equation (39) does not satisfy locality
condition but that it satisfies nonlocality condition. Concretely,
change of the population during a cycle depends of the population
at the end of this cycle. For this reason equation (39) represents
neither  the equation (25) nor equation (29) in which $k$ is
changed by $-k$.

In this way it is shown that, in distinction from set of all
"macroscopic" population dynamics (10), (12), set of all
"microscopic" population dynamics (21)-(24) (based on the
correspondence, biological discretization and locality condition)
is $T$  asymmetric. Or here $T$ applied on (21) does not yield
(22) and $T$ applied at (22) does not yield (21).

Now, on the basis of the previous discussions, it is not hard to
see that by "microscopic" analysis $C_{E}T$  symmetry is broken,
i.e. that neither (21) nor (22) is $C_{E}T$  invariant
(symmetric). Simply speaking here $T$ generates an ecologically
inplausible nonlocality that cannot be trenaformed in ecologically
plausible locality by $C_{E}$. Of course, even by "microscopic"
analysis (21) and (22) are $P$ invariant (symmetric), but neither
(21) nor (22) is $C_{E}PT$ invariant (symmetric).

In this way it is shown that $C_{E}PT$ symmetry of the unlimited
exponential population growth or decrease is explicitly broken by
"microscopic" analysis. Moreover, it is not hard to see that the
same conclusion on the explicit breaking of the $C_{E}PT$ symmetry
by "microscopic" analysis can be done for all other simple
population dynamical equations.

 \section {Conclusion}

In conclusion the following  can  be repeated and pointed out. In
ecology there is a $C_{E}PT$ symmetry theorem. It states that all
simple ecological, i.e. population dynamical equations (unlimited
exponential population  growth (or decrease) dynamics, logistic
or  Verhulst equation,  usual and generalized Lotka-Volterra
equations ) are "macroscopically" invariant (symmetric) in respect
to successive application of the time reversal transformation,
space coordinates reversal or parity transformation  and
predator-prey reversal transformation. However, such $C_{E}PT$
symmetry becomes explicitly broken by more accurate, "microscopic"
analysis .

\section {References}

\begin {itemize}

\item[[1]]  N.N.Bogolubov, A.A.Logunov, T.Todorov, {\it Introduction to Axiomatic Quantum Field Theory} (W.A.Benjamin, Reading, Mass., 1975.)
\item [[2]]  R.F.Streater, A.S.Wightman, {\it PCT, Spin and Statistics and All That} (Addison-Wesley, New York, 1989.)
\item [[3]]  H.T.Davies, {\it Introduction to Nonlinear Differential and Integral Equations} (Dover, New York, 1962.)
\item [[4]]  R.M.May, {\it Stability and Competition in Model Ecosystems}  (Princeton Univ.Press., Princeton, New Jersey,1974.)
\item [[5]] E .C.Pielou, {\it An Introduction to Mathematical Ecology} (John Wiley and Sons, New York, 1969.)
\item [[6]]  E.C.Pielou, {\it Mathematical Ecology} (John Wiley and Sons, New York, 1977.)

\end {itemize}

\end {document}